\documentclass[runningheads]{llncs}
\usepackage[T1]{fontenc}
\usepackage{cite}
\usepackage{amsmath,amssymb,amsfonts}
\usepackage{graphicx}
\usepackage{textcomp}
\usepackage{xcolor}
\usepackage{overpic}
\usepackage{todonotes}
\usepackage{mathtools}
\usepackage{hyperref}
\usepackage{xspace}
\usepackage{pgfplots}

\usepackage{algorithm}
\usepackage{algpseudocode}
\pgfplotsset{compat=1.18}

\newcommand{\cmnt}[1]{}
\newcommand{\ignore}[1]{}
\newcommand{\remove}[1]{}

\newcommand{\secref}[1]{Section~\ref{sec:#1}}
\newcommand{\figref}[1]{Fig.~\ref{fig:#1}}
\newcommand{\tabref}[1]{Table~\ref{tab:#1}}

\newcommand{\algoref}[1]{{Algorithm \ref{alg:#1}}}

\newcommand{\td} {TD\xspace}

\algrenewcommand{\algorithmiccomment}[1]{// #1}

\usepackage{graphicx}
\begin{document}
\title{Fault-Tolerant Decentralized Distributed Asynchronous Federated Learning with Adaptive Termination Detection}

\titlerunning{Fault-Tolerant Federated Learning}

\author{Phani Sahasra Akkinepally\inst{1} \and
        Manaswini Piduguralla\inst{1} \and
        Sushant Joshi\inst{1} \and
        Sathya Peri\inst{1} \and
        Sandeep Kulkarni\inst{2}}

\authorrunning{Akkinepally et al.}

\institute{Indian Institute of Technology Hyderabad, India \\
\and
Michigan State University, USA \\
}
\maketitle              
\begin{abstract}
Federated Learning (FL) facilitates collaborative model training across distributed clients while ensuring data privacy. Traditionally, FL relies on a centralized server to coordinate learning, which creates bottlenecks and a single point of failure. Decentralized FL architectures eliminate the need for a central server and can operate in either synchronous or asynchronous modes. Synchronous FL requires all clients to compute updates and wait for one another before aggregation, guaranteeing consistency but often suffering from delays due to slower participants. Asynchronous FL addresses this by allowing clients to update independently, offering better scalability and responsiveness in heterogeneous environments.

Our research\footnote{Code can be anonymously viewed here: \href{https://anonymous.4open.science/r/P2PFedMLCodes-3C32}{\textit{[Link]}}} develops an asynchronous decentralized FL approach in two progressive phases. (a) In Phase 1, we develop an asynchronous FL framework that enables clients to learn and update independently, removing the need for strict synchronization. (b) In Phase 2, we extend this framework with fault tolerance mechanisms to handle client failures and message drops, ensuring robust performance even under unpredictable conditions. As a central contribution, we propose \emph{Client-Confident Convergence} and \emph{Client-Responsive Termination} novel techniques that provide each client with the ability to autonomously determine appropriate termination points. These methods ensure that all active clients conclude meaningfully and efficiently, maintaining reliable convergence despite the challenges of asynchronous communication and faults.

\keywords{Federated Learning  \and Decentralized Framework  \and Asynchronous System  \and Crash Tolerance  \and Termination Detection}
\end{abstract}
\section{Introduction}
\label{sec:intro}
\begin{figure}[t]
    \centering
    \begin{overpic}[width=0.9\columnwidth]{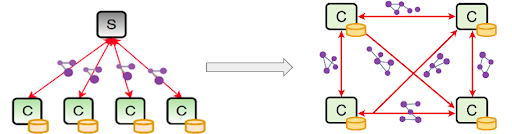}
        \put(20,-2){\textbf{(a)}}
        \put(77,-2){\textbf{(b)}}
    \end{overpic}
    \caption{Centralized to decentralized federated learning architecture. (a) Centralized approach with central server. (b) Decentralized peer-to-peer approach.}
    \label{fig:central_to_decentral}
    \vspace{-2em}
\end{figure}

Federated learning is a branch of Distributed Machine Learning (DML), in which nodes collaborate without sharing raw data. Clients conduct local model training and share parameters for aggregation, effectively preserving privacy. Centralized FL relies on a central server to coordinate client communication through two-way interactions between the server and each client, as shown in Figure~\ref{fig:central_to_decentral}(a). In contrast, decentralized FL uses peer-to-peer communication among clients, improving resilience, as illustrated in Figure~\ref{fig:central_to_decentral}(b). 

Traditionally, FL is implemented using a centralized setup, as described earlier. While conceptually straightforward, centralized FL inherently suffers from scalability issues and is vulnerable to single points of failure due to its reliance on a central coordinating server. A natural alternative is decentralized synchronous FL, where clients collaboratively train without relying on a central coordinator.




However, even in controlled environments such as machines connected in the same Local Area Network (LAN), message delay and relative speeds can make it difficult to assume synchronous mode of execution. Delays, packet reordering, and device variability lead to divergence and inconsistent progress across clients. Moreover, simulations often fail to accurately capture such real-world complexities, resulting in overly optimistic assumptions about system behavior. This realization motivates our exploration of multi-machine, multi-client real-world setups, allowing us to study asynchronous FL more rigorously under realistic conditions.

An equally critical aspect of real-world FL deployments is \emph{fault tolerance} which is the ability of the system to continue delivering correct results despite failures, such as clients dropping out or disconnecting. In this work, we specifically focus on benign crash faults, excluding more adversarial scenarios such as Byzantine faults~\cite{lamport1982byzantine}. 

In decentralized asynchronous setups, especially in multi-machine environments, there is no global notion of completion or termination of a work, known as \emph{termination detection} (\td). Termination detection in a distributed system \cite{tel2000distributed} is the problem of determining whether a distributed computation has completed—that is, all processes are idle and no messages are in transit. This can be challenging because each process only has partial information about the global system state, and local inactivity does not guarantee global completion. 

In distributed federated learning, the absence of a clear termination condition can lead to two significant issues: clients may either terminate prematurely before sufficient convergence is achieved, or continue training unnecessarily, wasting computational resources. To mitigate these risks, we introduce two key mechanisms: \emph{Client-Confident Convergence} and \emph{Client-Responsive Termination}.

The \emph{Client-Confident Convergence} mechanism ensures that a client initiates termination only after observing sufficient stability in the training process.
In response, the \emph{Client-Responsive Termination} mechanism ensures that any client receiving this signal updates its own termination flag and propagates the signal through its subsequent model broadcasts. This enables a distributed yet coordinated shutdown across all clients, improving robustness and avoiding indefinite training. (For detailed design, refer to \secref{frame}.)






In this work, we aim to contribute practical insights into designing federated learning systems that can work effectively (fault-tolerant) despite being the system being asynchronous and crash-prone. Specifically, in this work, we: (a) propose a fully decentralized convergence mechanism where each client autonomously monitors local model stability without centralized coordination in \secref{frame}.
(b) introduce a client-responsive termination protocol that ensures termination signals are reliably propagated, and decided when training had converged in \secref{frame}. (c) demonstrate the practical effectiveness of the approach through experiments in realistic, multi-machine federated learning environments in \secref{expts}.

\ignore{
\begin{itemize}
    \item Propose a fully decentralized convergence mechanism where each client autonomously monitors local model stability without centralized coordination in \secref{frame}.
    \item Introduce a client-responsive termination protocol that ensures termination signals are reliably propagated, and decided when training had converged in \secref{frame}.
    \item Demonstrate the practical effectiveness of the approach through experiments in realistic, multi-machine federated learning environments in \secref{expts}.
\end{itemize}
}


\section{Problem Statement}
\label{sec:prob}

The primary challenges addressed in this work stem from asynchronous nature of real-life networks and the issues that arise while handling crashes in asynchronous systems. Synchronous FL, though conceptually simpler to design, struggles in practical deployments where differences in device capabilities, intermittent connectivity, and varying computational resources lead to prolonged idle times and degraded system efficiency. These challenges are further amplified in decentralized, peer-to-peer setups, where the absence of a central coordinator makes achieving synchronization even more difficult.


Additionally, asynchronous decentralized FL introduces a critical challenge of termination. In the absence of global synchronization or a centralized authority, clients lack clear criteria to decide when to stop their participation. Without proper termination strategies, clients may either exit prematurely, compromising convergence, or continue unnecessarily, leading to wasted computation and communication overhead. Addressing this termination challenge is essential for building efficient and reliable FL systems, particularly in real-world, multi-machine deployments.

Handling crash failures in such an asynchronous setup further complicates the issues. Any real-world system is subject to crash failures. While byzantine failures ~\cite{lamport1982byzantine} are not uncommon in real-life systems, we focus on benign crash failures in this work. To tackle these issues comprehensively, our research is organized into two phases:

\vspace{1mm}
\noindent
\textbf{Phase 1: Addressing Asynchrony in Federated Learning}
In the first phase, we develop a fully asynchronous federated learning framework that enables clients to perform local model updates independently, without waiting for synchronization barriers. This design allows the system to naturally accommodate differences in client computation speeds and network delays. By removing the rigid requirement for synchronized rounds, we aim to achieve faster and more flexible convergence behavior, particularly in scenarios involving heterogeneous or resource-constrained devices.

\vspace{1mm}
\noindent
\textbf{Phase 2: Incorporating Fault Tolerance and Providing Termination Mechanisms}
While asynchronous models variability in participation and communication delays, real-world distributed systems are also subject to unexpected failures, including client crashes, disconnections, and intermittent communication losses. Therefore, in the second phase, we extend our asynchronous FL framework by integrating fault-tolerance mechanisms. These include strategies such as timeout-based failure detection, adaptive aggregation techniques, and distinguishing between slow and crashed clients.

Crucially, we also introduce a novel termination strategy tailored for asynchronous and fault-prone environments. This includes the \emph{Client-Confident Convergence} and \emph{Client-Responsive Termination} mechanisms, which empower clients to autonomously decide when to stop participating in the learning process based on local progress and system feedback. These mechanisms ensure that active clients exit training meaningfully, maintaining overall system efficiency and reliable convergence even in large-scale, decentralized deployments.

By structuring our research into these two phases, we aim to provide a comprehensive exploration of federated learning under practical deployment conditions, addressing not only the technical limitations of synchronization and fault tolerance but also solving the critical challenge of termination in asynchronous decentralized systems. This positions our work as a meaningful step toward enabling scalable, reliable, and efficient federated learning for real-world applications.

\vspace{-5mm}
\section{Proposed Framework}
\label{sec:frame}
\vspace{-2mm}
\subsection{System Model}

We consider a decentralized, federated machine learning (FedML) system comprising a set of $N$ clients, denoted as $C=\{c_1,c_2,\ldots,c_n\}$, where each client $c_i$ possesses their own local dataset $D_i$. These datasets are non-identical and independently distributed (non-IID) and may vary in size and distribution, reflecting practical heterogeneous data scenarios.
Unlike traditional centralized FL setups, our system does not rely on a central server for coordination. Instead, the clients are connected in a peer-to-peer (P2P) network topology where each client can communicate directly with all of the other clients. 

Each client maintains a local model and iteratively updates it using its local data. Periodically, clients exchange model updates with their neighbors through message passing. The model exchanges are in the form of model parameters. Messages sent over the network are assumed to be reliable. We assume no message loss, corruption, or duplication. However, communication delays are allowed and may vary between different client pairs.

\vspace{1mm}
\noindent
\textbf{Asynchronous Model:} The system follows a asynchronous communication model, where clients exchange messages over the network using sockets. Communication is fully decentralized, and the system is asynchronous, i.e., clients proceed with their local computations and communication at their speed without a common clock.


\vspace{1mm}
\noindent
\textbf{Failure Model:} The system assumes a crash fault model, where clients may become unresponsive due to software crashes, network disconnections, or resource limitations. These faults are benign—clients stop functioning without sending incorrect or malicious messages (i.e., no Byzantine behavior is considered). The model also supports temporary and intermittent failures, allowing clients to rejoin after transient faults. 


\subsection{Approach}
To systematically address the challenges identified in the traditional FL, this research is structured into two distinct phases, each tailored to tackle a specific dimension of complexity in Federated Learning (FL) systems namely, asynchronous behavior and fault tolerance. By adopting this phased methodology, we ensure that both the coordination challenges of decentralized systems and the resilience requirements of fault-prone environments are addressed in a gradual and controlled manner.

\begin{algorithm}
\caption{Synchronous Model}
\begin{algorithmic}[1]

\For{each client $C_i$}
    \State Initialize model $M_i$
    \State Set \texttt{$current\_round$} $\gets 0$
    \State Set \texttt{$previous\_round$} $\gets 0$
    
    \Comment{Repeat for $x$ rounds}
    \For{\texttt{$current\_round$} = 0 to $x$}
        \State Perform local model update on $M_i$

        \Comment{Broadcast model if round has progressed}
        \If{\texttt{$current\_round$} $\neq$ \texttt{$previous\_round$}}
            \State Broadcast $\langle M_i, \texttt{$current\_round$} \rangle$ to other $(n - 1)$ clients
        \EndIf

        \Comment{Wait to receive models from all other clients}
        \While{true}
            \If{received models from all other clients}
                \State Aggregate the received models and compute average $M_{\text{avg}}$
                \State Update $M_i \gets M_{\text{avg}}$
                \Comment{Mark round as complete}
                \State \texttt{$previous\_round$} $\gets$ \texttt{$current\_round$}
                \State Advance to next round
                \State \textbf{break}
            \EndIf
        \EndWhile
    \EndFor
\EndFor

\Comment{Handle received messages from other clients}
\If{a message $\langle M_j, r_j \rangle$ is received}
    \If{\texttt{$current\_round$} = $r_j$}
        \State Increment \texttt{$received\_count$} by 1
    \EndIf
\EndIf

\end{algorithmic}
\end{algorithm}

\vspace{1mm}
\noindent
\textbf{Phase 1: Managing Asynchronous Behavior Using Round-Based Coordination}

In the first phase, we focus on mitigating the effects of asynchronous inherent in distributed FL settings by leveraging a round-based synchronization strategy. Each client, during its communication phase, broadcasts its current round number along with its model updates to all other participating clients in the system. This exchange of round numbers provides a lightweight coordination mechanism, enabling clients to maintain awareness of the collective progress of the system despite differences in local update schedules or temporary communication delays.

This strategy not only facilitates smoother convergence of the global model but also prevents divergence due to inconsistent or out-of-order updates. To complement the synchronization mechanism, we implement a termination protocol, ensuring that all active clients reach a mutual agreement on when to halt training. This eliminates premature or inconsistent termination across the network, establishing a foundation for reliable learning outcomes in distributed environments.

\begin{algorithm}
\caption{Client Logic}
\begin{algorithmic}[1]
\Require ClientID \texttt{id}, ClientsList \texttt{$P = \{p_1, p_2, \cdots p_n$\}}

\State Initialize model, optimizer, train/test data loaders
\State Initialize tracking variables (rounds, weights, flags)

\While{$current\_round < R\_PRIME$}
    \Comment{Local Training}
    \For{$epoch \gets 1$ to $EPOCHS\_PER\_ROUND$}
        \State Train model on local data batch-wise
    \EndFor

    \State Extract model weights $localWeights$
    \If{termination flag received}
        \State Broadcast $localWeights$ with terminate flag
        \State \textbf{break}
    \EndIf

    \Comment{Broadcast and Wait}
    \State Broadcast $localWeights$ to peers
    \State Wait $TIMEOUT$ seconds for incoming weights

    \Comment{Crash Detection}
    \ForAll{peer $p$ in ClientsList}
        \If{no message from $p_i$ \textbf{and} $p_i$ not marked crashed}
            \State Mark $p_i$ as crashed
            \State Log the event
        \EndIf
    \EndFor

    \Comment{Model Update}
    \State Aggregate all received weights
    \State Update model with aggregated weights
    \State Evaluate accuracy on test set

    \Comment{Termination Criteria Check}
    \If{$current\_round \geq MINIMUM\_ROUNDS$}
        \If{$curr\_weight - prev\_weight > threshold$ \textbf{and} $recent\_crashes = None$}
            \State Increment convergence counter
        \Else
            \State Reset convergence counter
        \EndIf

        \If{$convergence\_counter \geq COUNT\_THRESHOLD$}
            \State Log termination condition met
            \State Broadcast $weights$ with terminate flag
            \State \textbf{break}
        \EndIf
    \EndIf

    \State Store current weights
    \State round++
    \State Clear message buffer
\EndWhile

\If{maximum rounds reached} \Comment{Termination Finalization}
    \State Log and broadcast final weights
\EndIf
\State Broadcast termination message to all peers

\end{algorithmic}
\label{alg:Exec}
\end{algorithm}

\vspace{1mm}
\noindent
\textbf{Phase 2: Achieving Fault Tolerance in Fully Asynchronous FL Systems}


The second phase of our research transitions to a fully asynchronous FL setup, where round-based coordination is removed, allowing clients to send updates entirely independently of one another.

To gracefully handle delays and failures, we introduce timeout-based mechanisms to detect potentially failed or unresponsive clients. In this approach, a client $C_i$ will wait for a message $m$ from another $C_j$ until the timeout expires. After which $C_i$ proceeds to the next round as shown in \algoref{Exec}. So, if $C_j$ had failed, then $C_i$ will not receive $m$. So, $C_i$ will proceed with the execution while marking $C_j$ as crashed. However, if the message $m$ is delayed then $C_i$ will consider $m$ in whatever round it receives and change the status of $C_j$ as alive. Unlike rigid synchronization, this approach allows the system to differentiate between slow and genuinely failed clients, minimizing unnecessary blocking and promoting progress even in degraded conditions.


To ensure effective learning despite faults, we propose the development of a client confidence-based convergence architecture and additionally, a client-responsive termination mechanism is incorporated to prevent indefinite waiting on failed or lagging clients. This enables the system to dynamically adapt to fluctuating participation levels while safeguarding convergence properties.


\noindent
\emph{Client-Confident Convergence:} It is a decentralized mechanism where each client independently executes the same termination logic during the federated learning process. Specifically, every client continuously monitors the progress of training by evaluating two key conditions: (a) The client has observed `$x$’ (convergence threshold) consecutive rounds without any detected crashes in the system. (b) The difference between the previous global model average and the current global model average falls below a predefined threshold, indicating diminishing model improvement.

When both conditions are satisfied, the client broadcasts a termination signal to all other active clients in the system. The rationale here is that the client encountering this stable state first has already witnessed $x$ stable rounds, suggesting that further training is unlikely to yield significant improvements. Importantly, this process is fully decentralized and can be triggered by any client in the system at runtime, making termination adaptive to dynamic network conditions and learning convergence.

\noindent
\emph{Client-Responsive Termination Protocol:} While signaling termination is essential, simply broadcasting a termination message is not sufficient in decentralized systems. Without a structured termination protocol, some clients might mistakenly interpret missing updates as client failures, leading to ambiguity regarding which clients have genuinely terminated versus those that might have crashed or disconnected.

To address this, we introduce the Client-Responsive Termination Protocol. In this mechanism, whenever a client receives a termination signal from another client, it updates its own internal termination flag in real time. From that point onward, the client continues participating in communication by broadcasting its model updates along with the termination flag enabled. This ensures that the termination signal propagates reliably throughout the network, even reaching clients that may have temporarily missed earlier signals due to delays or intermittent disconnections.

This approach adds an additional layer of robustness to the termination process, ensuring that even clients that were disconnected during the initial termination broadcast eventually receive the signal and terminate gracefully. By doing so, the system avoids false assumptions of crashes and maintains a clear, coordinated shutdown procedure across the distributed environment.
Through these two phases, our research presents a comprehensive approach to building federated learning systems that are not only capable of handling the natural asynchronous of decentralized computation but also robust enough to sustain learning in fault-prone, real-world environments. The pseudocode of the algorithm incorporating these techniques is shown in \algoref{Exec}.

\vspace{-5mm}
\section{Experiments and Analysis}
\label{sec:expts}
\vspace{-2mm}
To validate the effectiveness and robustness of the proposed federated learning framework, we conducted a comprehensive set of experiments under various simulated distributed learning conditions. These experiments were designed to assess system behavior in the presence of real-world challenges such as client speed, communication delays, and potential client failures.

\vspace{1mm}
\noindent
\textbf{Experimental setup:} The experimental evaluation was conducted with the number of clients ranging from 4 to 12, increasing in steps of 2. The system was developed in Python, leveraging threads to spawn individual client processes and sockets to enable communication between them in a fully decentralized, peer-to-peer manner. All experiments were executed on CPU-only environments. Two distinct phases were tested under different environmental assumptions. In Phase 1, the system operates through synchronization over an asynchronous system, with no client crashes permitted. In contrast, Phase 2 adopts an asynchronous setting where client crashes. The experiments were deployed across a multi-machine, multi-client setup, to simulate realistic distributed environments and test the robustness and scalability of the proposed approach.

\vspace{1mm}
\noindent
\textbf{System specifications:} The experiments were conducted using three high-performance machines in a distributed, multi-client, multi-machine environment. All machines communicated over a LAN (local area network), and experiments were executed entirely on CPUs without GPU acceleration as shown in \tabref{machine_specs}.

\begin{table}[h]
\centering
\begin{tabular}{|c|l|c|c|c|}
\hline
\textbf{Machine} &    \textbf{Operating} & \textbf{RAM} & \textbf{Physical} & \textbf{Clock} \\
  & \textbf{System} &     &    \textbf{Cores} & \textbf{Speed} \\
\hline
Machine 1 & Ubuntu 18.04.6 & 376 GiB & 56  & 4.0 GHz \\
Machine 2 & Ubuntu 22.04.5 & 251 GiB & 112 & 2.0 GHz \\
Machine 3 & Ubuntu 22.04.5 & 188 GiB & 52  & 3.5 GHz \\
\hline
\end{tabular}
\vspace{1 em}
\caption{System specifications of the machines used.}
\label{tab:machine_specs}
\vspace{-2em}
\end{table}

\vspace{1mm}
\noindent
\textbf{Data specifications:} Preliminary experiments were conducted using both the CIFAR-10 and MNIST datasets under IID (independent and identically distributed) and Non-IID conditions to establish baseline performance. However, the primary focus of this study is to analyze the behavior of federated learning systems under extreme and adverse conditions. As such, the subsequent experiments presented in this work specifically focus on Non-IID data distributions using the CIFAR-10 dataset, which better reflect real-world data heterogeneity and serve to stress-test the robustness of the proposed system. The dataset comprises 60,000 color images distributed across 10 classes, with 50,000 images used for training and 10,000 for testing. Each image has dimensions of $32 \times 32 \times 3$ pixels. To simulate Non-IID conditions, client-specific data partitions were generated using a Dirichlet distribution-based sampling strategy with $\alpha = 0.6$, which enables controlled variation in data skew by adjusting the concentration parameter.
The Convolutional Neural Network (CNN) model architecture used throughout the experiments comprises two convolutional layers followed by two fully connected layers, resulting in approximately 225,034 parameters (about 0.44 MB). Consequently, each communication round between clients involved the transmission of roughly $0.44$ MB of data for each client, representing model updates.


\vspace{1mm}
\noindent
\textbf{Phase 1 - Fault-Free System Experiments:} To establish baseline performance metrics, we initially evaluated single-client configuration to understand the lower bounds of our federated learning system. Table~\ref{tab:baseline_results} summarizes the classification accuracy achieved by individual clients under varying data distribution settings.

\begin{table}[h]
\centering
\caption{Baseline Performance Results}
\label{tab:baseline_results}
\begin{tabular}{|l|c|}
\hline
\textbf{Scenario} & \textbf{Accuracy (\%)} \\
\hline
Non-IID Single Client (Fixed data chunk) & 26.23 \\
IID Single Client (Fixed data chunk) & 37.48 \\
Single Client (Full dataset) & 70.82 \\
\hline
\end{tabular}
\vspace{-2em}
\end{table}

To understand the impact of data distribution and collaboration in federated learning, we evaluated three single-client training configuration.

In the first case (Non-IID Sgl), each client was assigned a fixed chunk of $5000$ data points drawn from a highly \textit{Non-IID distribution}, representing real-world data heterogeneity. These clients trained independently without any communication, and the average accuracy achieved across all clients was \textbf{26.23\%}, highlighting the limitations of learning from skewed, isolated data.
In the second scenario (IID Sgl), each client again received a fixed $5000$ data point chunk, but this time the data was sampled in an \textit{IID manner}, ensuring a balanced and representative distribution. This improved the average accuracy to \textbf{37.48\%}, demonstrating that even without collaboration, better data diversity significantly enhances performance.
The third case (Sgl Full) represents the \textit{ideal baseline}, where a single client is given access to the \textbf{entire training dataset}. Without any distribution or communication overhead, this setup yielded an accuracy of \textbf{70.82\%}, indicating the upper-bound performance that could be achieved in a centralized setting.

These results collectively emphasize the importance of collaboration in federated learning, particularly under Non-IID conditions, where isolated training yields significantly suboptimal outcomes.

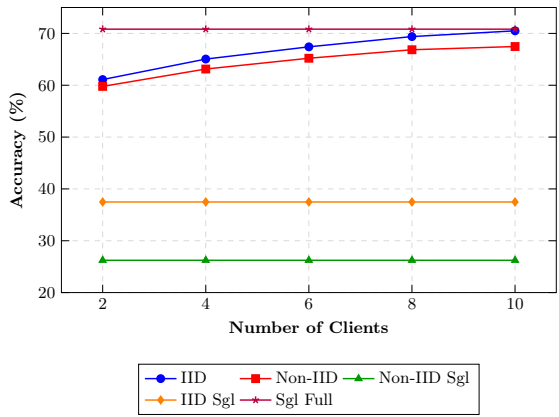
\begin{figure}
\centering
\begin{tikzpicture}[scale=0.7]
\begin{axis}[
    xlabel={\textbf{Number of Clients}},
    ylabel={\textbf{Accuracy (\%)}},
    grid=major,
    grid style={dashed, gray!30},
    legend style={
        font=\small,
        cells={align=left},
        at={(0.5,-0.25)},
        anchor=north
    },
    legend columns=3,
    width=0.9\linewidth,
    height=7cm,
    xtick={2,4,6,8,10},
    ymin=20, ymax=75,
    tick style={black},
    tick label style={font=\small},
    label style={font=\bfseries\small},
    legend cell align={left},
    mark options={solid}
]

\addplot[color=blue, mark=*, thick] coordinates {
    (2,61.1) (4,65.05) (6,67.41) (8,69.39) (10,70.5)
};
\addlegendentry{IID}

\addplot[color=red, mark=square*, thick] coordinates {
    (2,59.78) (4,63.13) (6,65.21) (8,66.85) (10,67.47)
};
\addlegendentry{Non-IID}

\addplot[color=green!60!black, mark=triangle*, thick] coordinates {
    (2,26.23) (4,26.23) (6,26.23) (8,26.23) (10,26.23)
};
\addlegendentry{Non-IID Sgl}

\addplot[color=orange, mark=diamond*, thick] coordinates {
    (2,37.48) (4,37.48) (6,37.48) (8,37.48) (10,37.48)
};
\addlegendentry{IID Sgl}

\addplot[color=purple, mark=star, thick] coordinates {
    (2,70.82) (4,70.82) (6,70.82) (8,70.82) (10,70.82)
};
\addlegendentry{Sgl Full}

\end{axis}
\end{tikzpicture}
\caption{Accuracy vs. Number of Clients for IID and Non-IID CIFAR-10 Settings with Single Client Baselines}
\label{fig:final_graph}
\end{figure}

\vspace{1mm}
\noindent
\emph{Results on CIFAR-10 Dataset:} The results in Figure~\ref{fig:final_graph} validate the effectiveness of the round-based synchronization in Phase 1, demonstrating proper coordination among clients.

As the number of clients increases from 2 to 10, accuracy steadily improves due to the inclusion of more data in each experiment. For Non-IID data, the accuracy rises from 59.78\% to 67.47\%, while for IID data it increases from 61.10\% to 70.50\%. 

\begin{table}[ht]
\centering
\begin{tabular}{|c|c|c|c|c|}
\hline
\textbf{Clients} & \textbf{Rounds} & \textbf{Accuracy (\%)} & \textbf{M1 Time (s)} & \textbf{M2 Time (s)} \\
\hline
2 & 31 & 59.78 & 172.19 & 170.81 \\
4 & 49 & 63.13 & 563.66 & 562.28 \\
6 & 62 & 65.21 & 2309.49 & 2309.52 \\
8 & 70 & 66.85 & 3720.45 & 3718.78 \\
10 & 80 & 67.47 & 3895.06 & 3882.56 \\
\hline
\end{tabular}
\caption{Non-IID CIFAR-10 Results}
\label{tab:non_iid}
\end{table}

\begin{table}[ht]
\centering
\begin{tabular}{|c|c|c|c|c|}
\hline
\textbf{Clients} & \textbf{Rounds} & \textbf{Accuracy (\%)} & \textbf{M1 Time (s)} & \textbf{M2 Time (s)} \\
\hline
2 & 29 & 61.10 & 157.31 & 154.75 \\
4 & 39 & 65.05 & 454.91 & 454.28 \\
6 & 45 & 67.41 & 1740.54 & 1740.54 \\
8 & 58 & 69.39 & 2000.24 & 2000.95 \\
10 & 65 & 70.50 & 3195.24 & 3204.25 \\
\hline
\end{tabular}
\caption{IID CIFAR-10 Results}
\label{tab:iid}
\end{table}

The system achieved consistent convergence with perfect synchronization across all runs. Termination was reliably detected using the designed consensus mechanisms. The results reveal the following insights:
\begin{itemize}
\item The IID scenario consistently achieves higher accuracy compared to non-IID, with a maximum accuracy of 70.50\% versus 67.47\%.
\item Both IID and non-IID scenarios demonstrate scalability, with accuracy improving as the number of clients increases.
\end{itemize}

These results provide a strong baseline for evaluating the fault-tolerant mechanisms implemented in Phase 2.

\vspace{1mm}
\noindent
\textbf{Phase 2: Fault Tolerance Experiments}
To systematically analyze the impact of client failures on federated learning, a series of targeted experiments were conducted under varying fault conditions. These experiments were designed to capture both the \textit{progressive degradation} and the \textit{scalability limits} of the system in the presence of crashes.To simulate a realistic system, we tested with three machine as shown in \tabref{machine_specs} with these machines were communicating with each other through the network. The 12 clients equally distributed among these machines. In experiments with lesser number of machines (such as 1 or 2), the clients were accordingly distributed.

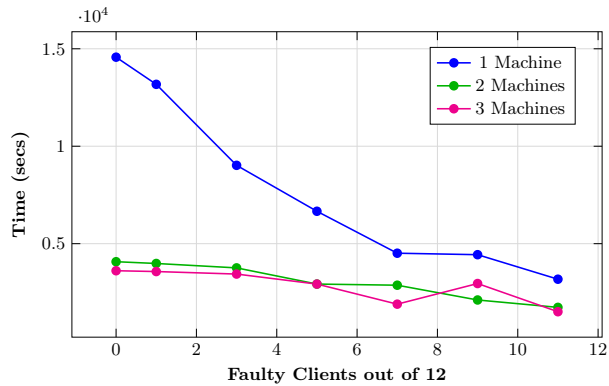
\begin{figure}
\centering
\begin{tikzpicture}[scale=0.75]
\begin{axis}[
    xlabel={\textbf{Faulty Clients out of 12}},
    ylabel={\textbf{Time (secs)}},
    grid=both,
    grid style={solid,gray!30},
    legend style={
        font=\small,
        at={(0.95,0.95)},
        anchor=north east
    },
    width=0.9\linewidth,
    height=7cm,
    tick style={black},
    tick label style={font=\small},
    label style={font=\bfseries\small},
    mark options={solid}
]

\addplot[color=blue, mark=*, thick] coordinates {
    (0,14569) (1,13176) (3,9020) (5,6663) (7,4511) (9,4435) (11,3177)
};
\addlegendentry{1 Machine}

\addplot[color=green!70!black, mark=*, thick] coordinates {
    (0,4075) (1,3984) (3,3758) (5,2928) (7,2869) (9,2111) (11,1732)
};
\addlegendentry{2 Machines}

\addplot[color=magenta, mark=*, thick] coordinates {
    (0,3612) (1,3570) (3,3441) (5,2927) (7,1897) (9,2955) (11,1509)
};
\addlegendentry{3 Machines}

\end{axis}
\end{tikzpicture}
\caption{Time taken with 12 Clients under Variable Fault Conditions}
\label{fig:exp1_time}
\end{figure}

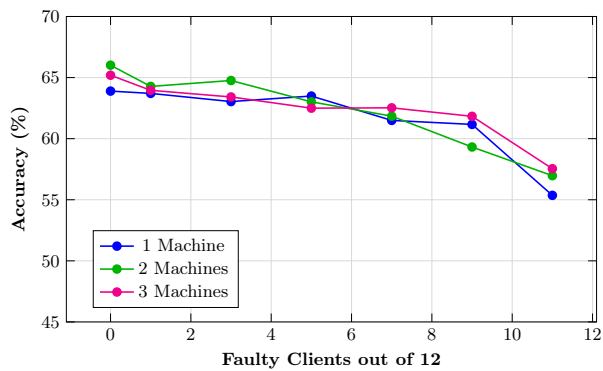
\begin{figure}
\centering
\begin{tikzpicture}[scale=0.75]
\begin{axis}[
    xlabel={\textbf{Faulty Clients out of 12}},
    ylabel={\textbf{Accuracy (\%)}},
    grid=both,
    grid style={solid,gray!30},
    legend style={
        font=\small,
        at={(0.05,0.3)},
        anchor=north west
    },
    width=0.9\linewidth,
    height=7cm,
    ymin=45, ymax=70,
    tick style={black},
    tick label style={font=\small},
    label style={font=\bfseries\small},
    mark options={solid}
]

\addplot[color=blue, mark=*, thick] coordinates {
    (0,63.8933) (1,63.7045) (3,63.0356) (5,63.4886) (7,61.4940) (9,61.1667) (11,55.36)
};
\addlegendentry{1 Machine}

\addplot[color=green!70!black, mark=*, thick] coordinates {
    (0,66.0142) (1,64.2764) (3,64.7611) (5,63.0143) (7,61.8420) (9,59.3167) (11,56.97)
};
\addlegendentry{2 Machines}

\addplot[color=magenta, mark=*, thick] coordinates {
    (0,65.1917) (1,63.9573) (3,63.4133) (5,62.4971) (7,62.5240) (9,61.83) (11,57.54)
};
\addlegendentry{3 Machines}

\end{axis}
\end{tikzpicture}
\caption{Performance with 12 Clients under Variable Fault Conditions}
\label{fig:exp1_acc}
\end{figure}

\noindent
\emph{Experiment 1: Variable Crash Analysis} evaluates system robustness by gradually increasing the number of crashed clients from 0 to 11 out of 12, revealing how performance metric, accuracy degrade with increasing fault intensity. \figref{exp1_acc} illustrates the variation in model accuracy with an increasing number of crash faults, while \figref{exp1_time} presents the total training time under the same fault conditions. Both figures compare the system behavior under single-machine, two-machine, and three-machine setups.

As expected, a gradual decline in accuracy is observed across all three configurations with increasing fault severity, demonstrating the system’s sensitivity to reduced client participation. However, Figure~\ref{fig:exp1_time} provides additional insight into the practical execution characteristics. In the case of zero faults, the single-machine setup incurs significantly higher training time compared to the multi-machine configurations. This elevated duration can be attributed to resource contention, since all 12 clients are hosted on a single physical machine, they compete heavily for CPU and memory, thereby introducing delays.


In contrast, when the same experiment is distributed across two and three machines, the total training time becomes more stable and comparable. This outcome emphasizes the importance of deploying federated systems in realistic, distributed environments rather than simulating all clients on a single machine. Additionally, involving machines with differing hardware specifications and clock speeds increases the system’s overall asynchrony, allowing for a more accurate evaluation of the framework’s performance and robustness in heterogeneous, real-world conditions.

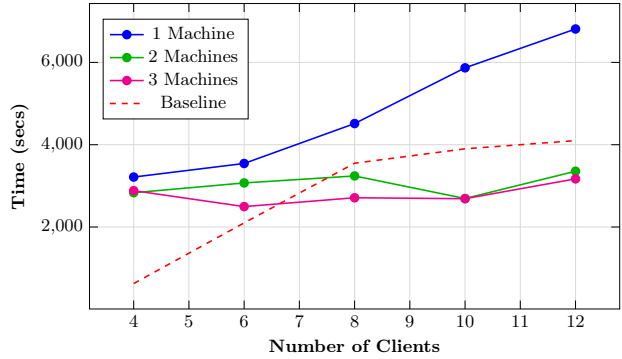
\begin{figure}
\centering
\begin{tikzpicture}[scale=0.75]
\begin{axis}[
    xlabel={\textbf{Number of Clients}},
    ylabel={\textbf{Time (secs)}},
    grid=both,
    grid style={solid,gray!30},
    legend style={
        font=\small,
        at={(0.3,0.95)},
        anchor=north east
    },
    width=0.9\linewidth,
    height=7cm,
    tick style={black},
    tick label style={font=\small},
    label style={font=\bfseries\small},
    mark options={solid}
]

\addplot[color=blue, mark=*, thick] coordinates {
    (4,3214) (6,3544) (8,4514) (10,5868) (12,6812)
};
\addlegendentry{1 Machine}

\addplot[color=green!70!black, mark=*, thick] coordinates {
    (4,2833) (6,3071) (8,3241) (10,2693) (12,3355)
};
\addlegendentry{2 Machines}

\addplot[color=magenta, mark=*, thick] coordinates {
    (4,2884) (6,2496) (8,2711) (10,2688) (12,3170)
};
\addlegendentry{3 Machines}

\addplot[color=red, dashed, thick] coordinates {
    (4,627) (6,2100) (8,3550) (10,3900) (12,4100)
};
\addlegendentry{Baseline}

\end{axis}
\end{tikzpicture}
\caption{Time taken with N/3 Faults and variable number of clients}
\label{fig:exp2_time}
\end{figure}

\begin{figure}
\centering
\begin{tikzpicture}[scale=0.75]
\begin{axis}[
    xlabel={\textbf{Number of Clients}},
    ylabel={\textbf{Accuracy (\%)}},
    grid=both,
    grid style={solid,gray!30},
    legend style={
        font=\small,
        at={(0.35,0.05)},
        anchor=south west
    },
    width=0.9\linewidth,
    height=7cm,
    tick style={black},
    tick label style={font=\small},
    label style={font=\bfseries\small},
    mark options={solid},
    ymin=50, ymax=70,
    xmin=4, xmax=12
]

\addplot[color=blue, mark=*, thick] coordinates {
    (4,62.03) (6,63.025) (8,65.90666667) (10,63.42) (12,62.53875)
};
\addlegendentry{1 Machine}

\addplot[color=green!70!black, mark=*, thick] coordinates {
    (4,58.85) (6,63.025) (8,65.57) (10,65.095) (12,63.35)
};
\addlegendentry{2 Machines}

\addplot[color=magenta, mark=*, thick] coordinates {
    (4,62.51) (6,63.025) (8,62.498) (10,64.33) (12,63.30714286)
};
\addlegendentry{3 Machines}

\addplot[color=red, dashed, thick] coordinates {
    (4,63.13) (6,65.21) (8,66.85) (10,67.47) (12,67.80)
};
\addlegendentry{Baseline}

\end{axis}
\end{tikzpicture}
\caption{Performance with N/3 Faults and variable number of clients}
\label{fig:exp2_acc}
\end{figure}
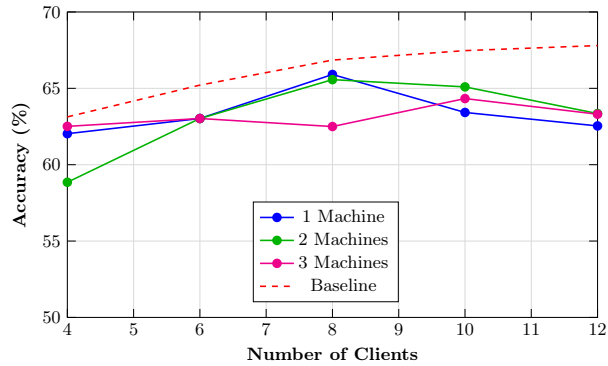

\noindent
\emph{Experiment 2: Proportional Fault Analysis} maintains a consistent failure rate (33\%) across different total client counts. Here the clients fail during the system execution at regular intervals. This setup helps assess whether the system can \textit{scale gracefully under consistent stress} and still converge effectively. These results are shown in Figures \ref{fig:exp2_time}, \ref{fig:exp2_acc}. Here the baseline refers to the non-faulty case executing with $\lfloor (2*n/3) \rfloor$ clients and are executing the learning algorithm outlined in phase1. When $n$ is 4 and 12, the actual number of clients in the baseline is 3 and 8 respectively. 


\figref{exp2_acc} illustrates that the three variants - single machine, two machine, and three machine setups. The experiment shows that the accuracy in case of the faulty environment is comparable to the baseline fault-free case even when handling $n/3$ client failures in an asynchronous environment. This showcases the robustness and adaptability of the proposed approach under partial client participation and non-deterministic execution patterns.

\figref{exp2_time} highlights the time-related impact of these configurations. While the single-machine setup exhibits a noticeable increase in training time due to higher resource contention, both the two-machine and three-machine setups manage to outperform the baseline in terms of computation time. This is because the client failures in these system are configured to occur sometime in the middle of the system execution and the failed clients help with the learning process till they fail. While this is not the case with the baseline which only has lesser number of clients throughout the execution. 


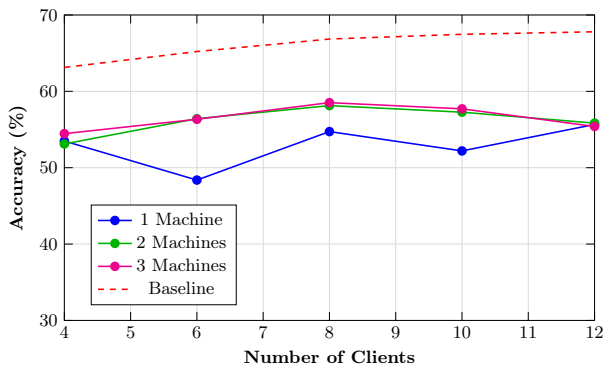
\begin{figure}
\centering
\begin{tikzpicture}[scale=0.75]
\begin{axis}[
    xlabel={\textbf{Number of Clients}},
    ylabel={\textbf{Accuracy (\%)}},
    grid=both,
    grid style={solid,gray!30},
    legend style={
        font=\small,
        at={(0.05,0.05)},
        anchor=south west
    },
    width=0.9\linewidth,
    height=7cm,
    tick style={black},
    tick label style={font=\small},
    label style={font=\bfseries\small},
    mark options={solid},
    ymin=30, ymax=70,
    xmin=4, xmax=12
]

\addplot[color=blue, mark=*, thick] coordinates {
    (4,53.49) (6,48.38) (8,54.73) (10,52.20) (12,55.66)
};
\addlegendentry{1 Machine}

\addplot[color=green!70!black, mark=*, thick] coordinates {
    (4,53.11) (6,56.42) (8,58.13) (10,57.27) (12,55.84)
};
\addlegendentry{2 Machines}

\addplot[color=magenta, mark=*, thick] coordinates {
    (4,54.46) (6,56.34) (8,58.52) (10,57.71) (12,55.40)
};
\addlegendentry{3 Machines}

\addplot[color=red, dashed, thick] coordinates {
    (4,63.13) (6,65.21) (8,66.85) (10,67.47) (12,67.80)
};
\addlegendentry{Baseline}

\end{axis}
\end{tikzpicture}
\caption{Performance with variable number of clients}
\label{fig:exp3_acc}
\end{figure}

For example, in the case of 12 clients, an $n/3$ failure corresponds to 4 client crashes, leaving 8 functional clients in the system. When comparing with a fault-free setup involving only 8 clients (the baseline), we observe that the time taken by the baseline is higher than the time required by the system operating with 12 clients and 4 faults. This clearly highlights the efficiency of the proposed asynchronous and fault-tolerant design, which is capable of leveraging partial participation more effectively than an asynchronous fault-free configuration with fewer active clients.

\noindent
\emph{Experiment 3: Maximum Fault Analysis} represents the \textit{worst-case scenario}, where only one client remains active. This setup allows for evaluating the system’s capability to tolerate extreme isolation and still make learning progress under minimal participation.

\begin{figure}
\centering
\begin{tikzpicture}[scale=0.75]
\begin{axis}[
    xlabel={\textbf{Number of Clients}},
    ylabel={\textbf{Time (secs)}},
    grid=both,
    grid style={solid,gray!30},
    legend style={
        font=\small,
        at={(0.05,0.95)},
        anchor=north west
    },
    width=0.9\linewidth,
    height=7cm,
    tick style={black},
    tick label style={font=\small},
    label style={font=\bfseries\small},
    mark options={solid},
    ymin=500, ymax=4500,
    xmin=3, xmax=13
]

\addplot[color=blue, mark=*, thick] coordinates {
    (4,1639) (6,1791) (8,2028) (10,2692) (12,3176)
};
\addlegendentry{1 Machine}

\addplot[color=green!70!black, mark=*, thick] coordinates {
    (4,1422) (6,1496) (8,1631) (10,1705) (12,1749)
};
\addlegendentry{2 Machines}

\addplot[color=magenta, mark=*, thick] coordinates {
    (4,1410) (6,1467) (8,1532) (10,1577) (12,1608)
};
\addlegendentry{3 Machines}

\addplot[color=red, dashed, thick] coordinates {
    (4,563.66) (6,2309.52) (8,3720.45) (10,3895.06) (12,4166.87)
};
\addlegendentry{Baseline}

\end{axis}
\end{tikzpicture}
\caption{Time taken with N-1 Faults and variable number of clients}
\label{fig:exp3_time}
\end{figure}
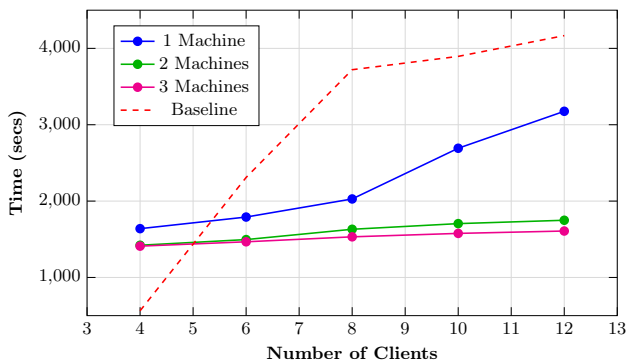

As shown in Figure~\ref{fig:exp3_acc}, the accuracy observed in the presence of $n-1$ faults is significantly lower than that of the synchronized no-fault system, which is expected due to the lack of client diversity and aggregation. However, it is noteworthy that the performance is still superior to the baseline case of a single client training independently on a fixed chunk of Non-IID data, as reported in Table~\ref{tab:baseline_results}. This highlights the benefit of collaborative learning—even with limited participation—over isolated learning on skewed datasets.
When considering the time behavior shown in Figure~\ref{fig:exp3_time}, we observe a natural decline in total training duration. This reduction is attributed to the smaller number of participating clients, leading to fewer communication rounds and significantly reduced coordination overhead.


\vspace{1mm}
\noindent
\textbf{Observations from the experiments:} The three fault experiments collectively validate the robustness, adaptability, and practical efficiency of our proposed asynchronous, fault-tolerant federated learning framework. 

\textbf{Experiment 1} demonstrated the system's graceful degradation under increasing crash faults. Even as the number of failed clients rose from 0 to 11, the system maintained reasonable accuracy and convergence time, particularly when deployed in a multi-machine setup. This highlighted the effectiveness of distributing clients across machines to reduce contention and better reflect real-world asynchronous.

\textbf{Experiment 2} examined the system’s behavior under a consistent failure rate of $n/3$ across varying client counts. The results showed that despite substantial asynchronous and partial failures, the proposed approach performed comparably to a fault-free baseline. This confirmed that the system can scale under fixed-stress conditions without significant accuracy or efficiency loss.

\textbf{Experiment 3} tested the extreme boundary of the system—when only one client remains active. While the accuracy dropped significantly, as expected, it still outperformed the Non-IID single-client baseline (Table~\ref{tab:baseline_results}), underlining the benefit of initial collaborative updates and the residual value of even limited federation.





\section{Related Work}
\label{sec:related}
Federated Learning was first introduced as a practical method for training machine learning models across distributed devices while preserving data privacy~\cite{mcmahan2017communication}. The seminal work by McMahan et al. ~\cite{mcmahan2017communication} presented \textbf{Federated Averaging (FedAvg)}, which operates through iterative model averaging where clients perform local stochastic gradient descent updates followed by server-side aggregation. FedAvg demonstrated significant communication efficiency, reducing required communication rounds by 10-100x compared to synchronized stochastic gradient descent.


The theoretical foundations of FedAvg convergence have been extensively studied under various conditions~\cite{sun2022decentralized,wang2021novel}. Wang and Ji ~\cite{wang2022unified} provided a unified convergence analysis for federated learning with arbitrary client participation, introducing a generalized version of federated averaging that amplifies parameter updates at intervals of multiple FL rounds. Their analysis captures the effect of client participation in a single term, obtaining convergence upper bounds for both non-stochastic and stochastic participation patterns. Li et al.~\cite{wang2021novel} established convergence rates for strongly convex and smooth problems in non-IID data settings, revealing that data heterogeneity slows convergence and necessitates decaying learning rates. Recent work has challenged pessimistic theoretical predictions, showing that FedAvg can achieve identical convergence rates in both homogeneous and heterogeneous data settings under certain conditions~\cite{beikmohammadi2024convergence}.




\vspace{1mm}
\noindent
\textbf{Synchronous Centralized Federated Learning Systems}

Synchronous federated learning has been extensively studied from system optimization perspectives, with comprehensive surveys identifying key bottlenecks in client selection, configuration, and reporting phases~\cite{survey2023acm}. Communication-efficient approaches for synchronous FL have focused on adaptive aggregation strategies in client-edge-cloud architectures~\cite{luo2024communication}. Luo et al.~\cite{luo2024communication} propose theoretical convergence analysis under aggregation frequency control, enabling dynamic resource allocation while maintaining model convergence guarantees. Their FedAda method demonstrates up to 4\% improvement in test accuracy, 6.8× shorter training time and 3.3× less communication overhead compared to prior solutions.

Feng et al.~\cite{feng2024understanding,feng2024towards} studied federated learning efficiency over wireless networks, deriving analytical expressions to characterize FL convergence rates accounting for transmission reliability, scheduling policies, and momentum methods. Their analysis reveals that delicately designed user scheduling policies or expanding bandwidth can expedite model training in reliable networks, but these methods become ineffective when connections are erratic. The incorporation of momentum method into model training algorithms accelerates convergence rate and provides greater resilience against transmission failures~\cite{feng2024towards}.

\vspace{1mm}
\noindent
\textbf{Real-World Asynchrony in Federated Learning}

The transition from theoretical synchronous models to practical asynchronous implementations addresses critical real-world challenges including device heterogeneity, intermittent connectivity, and varying computational resources~\cite{liu2021adaptive,async2024survey}. Liu et al.~\cite{liu2021adaptive} propose an adaptive asynchronous federated learning (AAFL) mechanism for resource-constrained edge computing, where a certain fraction of local updates are aggregated by arrival order at the parameter server in each epoch. 


Asynchronous federated learning eliminates the need for synchronized communication, allowing devices to contribute updates at their own pace~\cite{async2024survey}. Recent advances focus on addressing efficiency challenges through prospective momentum aggregation and fine-grained correction techniques~\cite{forootani2024asynchronous}. The work by Liao et al.~\cite{AsynFL:Liao+:ACM:2024} contributes to improved robustness in federated learning through decentralization and asynchronous updates. While this approach effectively addresses device heterogeneity, it does not explicitly implement fault tolerance mechanisms, leaving gaps in handling system failures during training. Similarly, MPLS \cite{Xu+:MPLS:Europar:2025} is a decentralized federated learning method that speeds up training by letting devices share and combine important parts of their models asynchronously, helping handle different device capabilities in edge environments; however, it does not address fault tolerance in the system.

Morell et al.~\cite{MORELL+:FGCS:2022} introduces a dynamic and adaptive fault-tolerant asynchronous federated learning framework utilizing volunteer edge devices. While this demonstrates the feasibility of distributed training across platforms like web browsers and terminal processes, such proof-of-concept environments fail to fully capture the computational heterogeneity, network variability, and rich data distributions encountered in real-world deployments. This limits their practical applicability for evaluating fault-tolerant learning at scale. A centralized approach with sequential model combination using first-come-first-serve aggregation is introduced by Ma et al.~\cite{realTimeFL:Ma+:IEEE:2021}. This work demonstrates practical applications of asynchronous FL in industrial fault diagnosis scenarios, though its centralized nature may present single points of failure.The potential of decentralized architectures to enhance system resilience over centralized alternatives has been extensively discussed in the literature~\cite{ieee2023survey}. However, existing approaches often suffer from high communication overhead and significant message complexity, which limit their practicality in real-time, fault-tolerant federated learning scenarios~\cite{ranellucci2022learning}. This highlights the pressing need for more efficient, scalable solutions that can deliver fault tolerance without compromising responsiveness.

The reviewed literature reveals a clear progression from early synchronous centralized approaches to more sophisticated asynchronous and decentralized methods~\cite{survey2023acm,ieee2023survey}. However, existing work focuses on robustness through decentralization without explicit fault tolerance or addresses fault tolerance with prohibitive computational overhead~\cite{ranellucci2022learning}. The challenge of achieving fault-tolerant federated learning in real time even with simple averaging mechanisms to start with, while maintaining efficiency remains largely unaddressed~\cite{liu2021adaptive,feng2024towards}. This gap motivates research on lightweight fault tolerance mechanisms that can operate effectively in distributed learning environments in real time without compromising the fundamental benefits of federated learning paradigms~\cite{luo2024communication,wang2022unified}.

\section{Conclusion}
\label{sec:Conclusion}
In this work, we presented a fully asynchronous, fault-tolerant federated learning framework designed to operate effectively under realistic deployment conditions characterized by client heterogeneity, network instability, and the absence of centralized control. Through systematic experimentation, we demonstrated the framework's robustness to crash faults, its ability to scale under partial failures, and its resilience even in extreme scenarios with minimal active clients. The proposed Client-Confident Convergence and Client-Responsive Termination mechanisms address a long-standing challenge in decentralized asynchronous FL achieving reliable, efficient, and autonomous termination without global coordination. By combining these contributions with practical fault-tolerance strategies, our framework enables scalable and efficient federated learning across distributed, failure-prone environments. 
In future we will focus on extending these guarantees to tolerate Byzantine faults, where clients may behave arbitrarily or maliciously due to software bugs, or adversarial manipulation. Incorporating Byzantine resilience is essential for deploying FL in open, untrusted environments such as cross-organizational collaborations or public networks

\bibliographystyle{splncs04} 
\bibliography{citations} 

\end{document}